\documentclass[aps,prd,twocolumn,groupedaddress,showpacs]{revtex4-1}

\usepackage[T1]{fontenc}

\usepackage{amsmath}
\usepackage{mathrsfs}
\usepackage{amssymb}
\usepackage{graphicx}
\usepackage{physics}

\usepackage{enumitem}
\usepackage{tikz}
\usepackage{circuitikz}
\usepackage{scalerel}
\usepackage[compat=1.0.0]{tikz-feynman}
\usepackage{epsfig}
\usepackage{subcaption}
\usepackage{float}

\usepackage{hyperref}
\hypersetup{
    colorlinks=true,
    linkcolor=blue,
    filecolor=magenta,      
    urlcolor=blue,
    citecolor=blue
}

\begin{document}

\title{Testing Higgs {\boldmath $CP$} properties at the CEPC with an additional ISR parameter}

\author{A. Drutskoy and E. Vasenin}

\affiliation{P.N. Lebedev Physical Institute of the Russian Academy of Sciences, Moscow 119991, Russia}

\date{\today}

\begin{abstract}

We evaluate the experimental sensitivity to the $CP$-odd admixture in the standard Higgs 
boson for the process $e^+e^-\to HZ$. The analysis is performed assuming the future lepton collider CEPC reference
detector operating at $\sqrt{s}=240~\text{GeV}$ with statistics of $5.6~\text{ab}^{-1}$.
Using the {\footnotesize WHIZARD}
generator with the Higgs Characterisation model and the {\footnotesize DELPHES} detector simulation framework we obtain data samples for 
different $CP$-odd Higgs admixture parameters $\widetilde{c}_{ZZ}$. The initial state radiation effects are taken into 
account in {\footnotesize WHIZARD}. We develop a novel data-analysis method that exploits the initial state radiation induced 
shift in the reconstructed event energy $E_{\text{RECO}}$ as an additional source of information on the Higgs $CP$ 
structure. The method combines the initial state radiation sensitive event-energy information with three 
angular observables in a multidimensional likelihood analysis. The expected upper limit on the $CP$-odd 
Higgs admixture is improved by 15\% compared with the standard analysis based on three angular variables.

\end{abstract}

\pacs{11.30.Er, 12.60.Fr, 13.66.Fg, 14.80.Bn}

\maketitle

\section{INTRODUCTION}

The discovery of the Higgs boson with a mass near $125~\mathrm{GeV}$
completed the particle content of the Standard Model (SM) 
and opened up a new frontier: searches for Beyond the Standard model (BSM) physics by means of
precise measurements of the Higgs boson parameters~\cite{201230}. One of these parameters
is the Higgs boson quantum numbers.
Within the SM the Higgs boson is predicted to be 
the $CP$-even scalar ($J^{CP}=0^+$). 
Any admixture of the $CP$-odd component will indicate BSM physics,
with deep implications for the baryogenesis 
and electroweak scale dynamics~\cite{2212.05833}. 

The Higgs boson is the excitation of the field responsible for
electroweak symmetry breaking and, through its gauge and Yukawa
interactions, for the masses of the weak gauge bosons and charged
fermions. Its measured mass and couplings determine the structure of
the electroweak sector, making precision tests of its quantum numbers
and interaction strengths an important test of the internal
consistency of the SM. At the same time, the Higgs
sector provides a sensitive portal to physics beyond the SM: extended
scalar sectors and higher-dimensional interactions may modify the
tensor structure of the Higgs couplings and introduce $CP$-odd
contributions.

Compared with hadron colliders, electron--positron colliders provide a
precisely known and tunable initial state, substantially smaller QCD
backgrounds, and a cleaner experimental environment. These features
enable model independent measurements based on the recoil mass
technique and precise reconstruction of Higgs production and decay
kinematics~\cite{cepc_whitemass}. The Circular Electron Positron Collider (CEPC)~\cite{cepc_a_tdr} is designed
to operate as a Higgs factory at a center of mass energy of
$\sqrt{s}=240~\mathrm{GeV}$. Its reference detector is optimized for
high-precision tracking, particle-flow reconstruction, and efficient
lepton identification, which are particularly important for the
$Z\to\mu^+\mu^-$ recoil channel considered in this work~\cite{cepc_ref_tdr}. 

Besides the CEPC~\cite{cepc_a_tdr}, several other proposed
electron--positron facilities, including the FCC-ee~\cite{fcc},
ILC~\cite{ilc_1,ilc_2,ilc_3}, and CLIC~\cite{clic_1}, are expected to
operate as Higgs factories. Precision measurements of the
Higgsstrahlung process $e^+e^-\to ZH$ and vector boson fusion
processes $e^+e^-\to H\ell\bar{\ell}$ provide high sensitivity to
possible $CP$-even and $CP$-odd contributions to the $HVV$ and
$Hf\bar f$ interactions. Here, $\ell=e,\mu$ denotes charged lepton
species, so that $\ell\bar{\ell}$ corresponds to either $e^-e^+$ or
$\mu^-\mu^+$; $V=W,Z$ denotes an electroweak vector boson; and $f$
denotes a fundamental SM fermion, i.e. a lepton or a quark.
Dedicated detector simulations demonstrate that a percent level 
or better precision on the $CP$-mixing parameters is 
achievable with a collected integrated luminosity 
of several $\text{ab}^{-1}$~\cite{ilc_higgs,1804.01241,chinese_cp}.

Experimentally, $CP$ information is extracted from observables that are odd under parity. 
In particular, the transverse 
spin correlations of the $\tau$ decay products 
in the decay $H\to\tau^+\tau^-$ and angular distributions in the
$e^+e^-\to ZH$ (with $Z\to \ell^+\ell^-$) process 
can be translated directly into differences between the $CP$-even and $CP$-odd couplings. Analysis 
technique using optimal observables 
and matrix element based methods have been developed and validated using
detailed detector simulation. 
These approaches both benefit from the low backgrounds and well defined 
initial state of the $e^+e^-$ collisions~\cite{2304.04390,1804.01241}. 

The precise lepton collider measurements and LHC (Large Hadron Collider) results at high energy can complement each other.
Important limits on the $CP$-odd admixtures are obtained by LHC. However, their sensitivity, especially to small $CP$-violating phases in fermionic couplings, 
can be significantly improved by Higgs factories measurements. 
With these measurements we can obtain model-independent evaluation of coupling tensors and phase-sensitive 
observables. Together, these experiments will either constrain the $CP$-violating parameters in
extensions of the Higgs sector or observe evidence of the BSM physics at the electroweak scale~\cite{2212.05833,2412.13130}.

Several studies have investigated upper limits on the $CP$ parameters in different scenarios. 
Recent ATLAS and CMS measurements have constrained anomalous $CP$-even
and $CP$-odd contributions to the $HVV$ interaction using the full
Run~2 data set. An ATLAS combination constrains the $CP$-odd SMEFT
coefficient $c_{H\widetilde W}$ to the interval
$[-0.14,0.49]$ at 95\% confidence level for
$\Lambda=1~\mathrm{TeV}$, when only the linear contribution of the
corresponding dimension-six operator is retained~\cite{atlas_prl}. Using $138~\mathrm{fb}^{-1}$ of data in the
$H\to\gamma\gamma$ channel, CMS obtained 95\% confidence
intervals
$f_{a3}\in[-1.5,\,1.5]\times10^{-4}$ and
$f_{a2}\in[-5.5,\,1.2]\times10^{-4}$~\cite{cms_cp}.
Here, $f_{a3}$ and $f_{a2}$ parameterize $CP$-odd and anomalous $CP$-even
$HVV$ tensor contributions, respectively.
In the paper~\cite{chinese_cp} the $HZZ$ tensor coupling $\widetilde{c}_{ZZ}$ is 
constrained to the region $[-0.06, 0.06]$ at $68\%$ CL for the future CEPC detector. ECFA collaboration studied 
sensitivity to $CP$-odd coupling in the $HZZ$ vertex for the future FCC-ee project~\cite{ecfa}. Their expected 
sensitivity is $f_{CP}^{HZZ} = \pm 1.2\times 10^{-5}$ at $68\%$ CL. The accurate $CP$-violating parameters description 
is given below in the theoretical part. 
All these studies use
likelihood method with three angular observables $\phi, \theta_1, \theta_2$, described below. To improve it, we develop a 
novel ISR-assisted data-analysis method that uses the ISR-induced shift in the reconstructed total event energy 
$E_{\text{RECO}}$ as an additional $CP$-sensitive observable.

Current experimental upper limits on the electron and neutron electric dipole moments 
(EDMs) can constrain possible $CP$-odd couplings of the Standard Model Higgs boson within 
specific BSM frameworks, such as the Two-Higgs-Doublet Model (2HDM)~\cite{edm_1}. However, these constraints are
strongly dependent on the BSM model and its internal parameters. Therefore, the calculation of EDM via a 
Barr–Zee-type two-loop diagram with a virtual Higgs boson and direct searches for the $CP$-odd component of the 
Higgs boson at colliders are complementary approaches. Under some conditions the EDM calculations can provide 
the limits on the same level or even better. More details can be found in the theoretical review~\cite{edm_2}.

Due to initial state radiation (ISR) effects, the reconstructed total event energy differs from 
the nominal center of mass energy. The total cross section of the processes with $HZZ$ vertex 
depends on the $CP$ properties of the coupling. Therefore, the $E_{\text{RECO}}$ distribution is 
sensitive to the $CP$ properties. Thus, rather than treating ISR solely as a radiative effect that 
smears the nominal collision energy, the developed method exploits it as an additional source 
of $CP$-sensitive information, effectively providing an energy scan below the nominal center of mass energy.
To our knowledge, this is the first Higgs $CP$ analysis to exploit ISR-induced event-energy 
variations as an explicit source of experimental sensitivity.

In this study, we focus on the process $e^+e^- \to ZH$ with the subsequent decay $Z\to \mu^+\mu^-$, which 
provides a clean and well reconstructed final state for 
probing the tensor structure of the $HZZ$ vertex at future Higgs factories. Operating at a center of mass 
energy of 240 GeV, the proposed Circular Electron Positron Collider 
(CEPC) can provide an optimal environment for precision studies of Higgs properties. Over one million Higgs 
bosons is expected with statistics of $5.6~\text{ab}^{-1}$ in a low background environment. The 
$e^+e^- \to ZH$ process provides an opportunity to accurately reconstruct production and decay angles and an angle
between the respective planes using precisely measured muons from the $Z$ boson decay. These angles 
are sensitive to possible $CP$-odd admixtures in the Higgs coupling to the $Z$ boson. We aim to 
assess the potential of the 
CEPC to constrain or reveal new sources of $CP$ violation in 
the electroweak symmetry breaking sector.

There are several equivalent ways to parameterize the tensor structure of the $HZZ$ interaction in an effective Lagrangian. 
In this work, we use the Higgs Characterisation (HC) framework~\cite{hc}, 
which provides a convenient and gauge-invariant description of possible $CP$-even and $CP$-odd admixtures in the Higgs sector. The 
effective Lagrangian for the interaction of scalar and pseudoscalar Higgs states with the $Z$ boson has the following form:
\begin{multline}
\mathcal{L}_{HZZ}
 = \frac{1}{2}\,\kappa_{\mathrm{SM}}\cos\psi_{CP}\, g_{HZZ}\, Z_\mu Z^\mu H - \\
 - \frac{1}{4\Lambda}
 \Big[
 \cos\psi_{CP}\,\kappa_{HZZ}\, Z_{\mu\nu} Z^{\mu\nu} + \\
 + \sin\psi_{CP}\,\kappa_{AZZ}\, Z_{\mu\nu}\tilde{Z}^{\mu\nu}
 \Big] H,
\label{eq:Lagrangian}
\end{multline}
where $Z_{\mu\nu}$ denotes the field-strength tensor and $\tilde{Z}^{\mu\nu}$ its dual. 
The parameter $\psi_{CP}$ is the $CP$-mixing angle, equal to zero in the SM, $\Lambda$ is the BSM physics energy scale. 
The first term in Eq.~(\ref{eq:Lagrangian}) corresponds to the SM $HZZ$ vertex, 
the second term represents a $CP$-even tensor contribution with coupling $\kappa_{HZZ}$, 
and the last term encodes a $CP$-odd tensor component proportional to $\kappa_{AZZ}$.

A more conventional notation introduces the parameter $\widetilde{c}_{ZZ}$, in terms of which 
the $CP$-violating part of the effective Lagrangian can be expressed as  
\begin{equation}
\mathcal{L}_{CPV} = \frac{H}{v} \left( \widetilde{c}_{ZZ} \, \frac{g_1^2 + g_2^2}{4} \, Z_{\mu\nu} \tilde{Z}^{\mu\nu} \right),
\end{equation}
where $g_1 = 0.358$ and $g_2 = 0.648$ are the electroweak gauge couplings, and $v$ 
denotes the vacuum expectation value of 
the Higgs field. The $\psi_{CP}$ values are subsequently converted to $\widetilde{c}_{ZZ}$ for 
the comparison with other analyses.

Another way to parameterize $CP$ violation is the $f_{CP}$, $f_{g2}$ and $f_{g4}$ notations described in 
detail in~\cite{atlas_hzz}. However, 
the $f$ parameters depend on the process cross section and 
the center of mass energy, 
therefore there are no direct conversion between $f_{CP}(f_{gi})$ and $\widetilde{c}_{ZZ}$. In case of experimental setup
in this study, the upper limits on the $CP$-violation parameters can be approximately related: 
\begin{equation}
    f_{CP} = 4\times10^{-3}\cdot\widetilde{c}_{ZZ}^2,
\end{equation}
which holds in the limit of small parameters. The similiar experimental setup is expected at the FCC-ee experiment, thus the upper limits
can be approximately converted from $f_{CP}^{HZZ} = 1.2\times 10^{-5}$ to $\widetilde{c}_{ZZ} \approx 0.06$, which was obtained assuming 
the statistics of $10.8~\text{ab}^{-1}$.

\section{Experimental procedures}

\subsection{Monte Carlo simulation and event reconstruction}

The \footnotesize WHIZARD \normalsize generator version 3.1.6~\cite{whizard_1,whizard_2} is used to calculate matrix elements and perform phase-space Monte Carlo simulations.  
The Higgs Characterisation model is included in the generator, and samples with different values of $\psi_{CP}$, corresponding to $\widetilde{c}_{ZZ}$ in the range $[-1.2, 1.2]$ are generated.  
The event generation is performed for the process
\[
e^+e^- \to HZ,\quad H \to \text{incl.},\quad Z \to \mu^+\mu^-
\]
at the center of mass energy $\sqrt{s} = 240~\text{GeV}$, taking into account ISR effects.
The generator \footnotesize WHIZARD \normalsize calculates ISR effects to all orders in perturbation theory
for the soft photon radiation and hard-collinear photons are calculated explicitly order by order, up to the third order in this analysis.
Potentially the accuracy of the ISR effects calculations can be tested using the process
$e^+e^- \to \mu^+\mu^- \gamma_{\text{ISR}}$ by comparing the $M(\mu^+\mu^-)$
distributions in data and MC simulations.
The parameter $\kappa_{HZZ}$ is set to zero, while $\kappa_{AZZ}$ and $\Lambda$ are chosen in such a way that the total cross section of the process remains 
independent of $\psi_{CP}$, ensuring that the total number of events in each sample is the same. The $CP$-odd admixture is then regulated by changing only the $CP$-mixing angle.

Subsequently, Pythia6~\cite{pythia} is used to simulate hadronization and final state radiation (FSR) effects.  
The resulting samples are then passed to the \footnotesize DELPHES \normalsize package~\cite{delphes} with the CEPC detector card for fast detector simulation.  

The $Z$ boson is reconstructed from the two muons.  
The Higgs boson is not reconstructed in this analysis.  
Its four-momentum is obtained as the recoil against the $Z$ boson, assuming two body final state.

\subsection{Angular distributions}

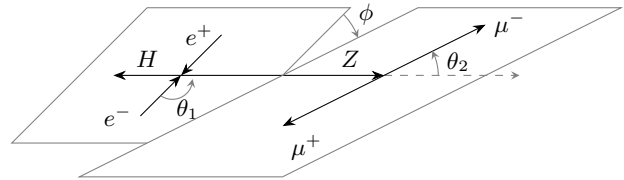
\begin{figure}[h]
    \centering
    \begin{tikzpicture}[scale = 0.9]

        \draw[-stealth, color=gray, dashed] (3.0,0) -- (5.0,0);
        \draw[-Stealth] (0,0) -- (3.0,0);
        \draw[] (2.5,0.25) node{$Z$};

        \draw[-Stealth] (0,0) -- (-1.0,0);
        \draw[] (-0.5, 0.25) node{$H$};

        \draw[-Stealth] (0.6,0.6) node[anchor = east] {$e^+$} -- (0,0);
        \draw[-Stealth] (-0.6,-0.6) node[anchor = east] {$e^-$} -- (0,0);

        \draw[-stealth, color = gray] (-0.3,-0.3) arc(235:360:0.3);
        \draw[] (0.1, -0.5) node{$\theta_1$};

        \draw[-Stealth] (3.0, 0) -- (4.5,0.75) node[anchor = west]{$\mu^-$};
        \draw[-Stealth] (3.0, 0) -- (1.5,-0.75) node[anchor = north west]{$\mu^+$};

        \draw[-stealth, color = gray] (3.8,0) arc (0:27:0.8);
        \draw[] (3.8, 0.2) node[anchor = west]{$\theta_2$};

        \draw[color = gray] (1.5, 0) -- (-1.5, -1.5) -- (1.5, -1.5) -- (6.5, 1.0) -- (3.5, 1.0) -- (1.5, 0);
        \draw[color = gray] (1.5, 0) -- (2.5,1.0) -- (-0.5, 1.0) -- (-2.5, -1.0) -- (-0.5, -1.0);

        \draw[-stealth, color = gray] (2.4, 0.9) arc (45:19:0.9);
        \draw[] (2.5, 0.9) node[anchor = west]{$\phi$};

    \end{tikzpicture}
    \caption{Kinematics for the $e^+e^-\to HZ,~Z \to \mu^+\mu^-$ process.}
    \label{fig:angles}
\end{figure}

Three angular observables, $\phi$, $\theta_1$, and $\theta_2$, are used in the analysis to study the 
$CP$ properties of the Higgs boson.  
The angle $\theta_1$ is defined as the production angle between the direction of the $e^-$ beam and 
that of the $Z$ boson.  
The angle $\theta_2$ is the decay angle between the $Z$ boson direction and the direction of the final state $\mu^-$ 
in the $Z$ boson rest frame.  
The angle $\phi$ is the angle between the $HZ$ production plane and the $Z$ decay plane.  
These three angles are shown in Fig.~\ref{fig:angles}.  

These observables are sensitive to the $CP$ properties of the Higgs boson and therefore exhibit different distributions for various admixtures of $CP$-odd and $CP$-even components.

\begin{figure*}
    \begin{subfigure}[t]{0.48\textwidth}
    \includegraphics[width = \textwidth]{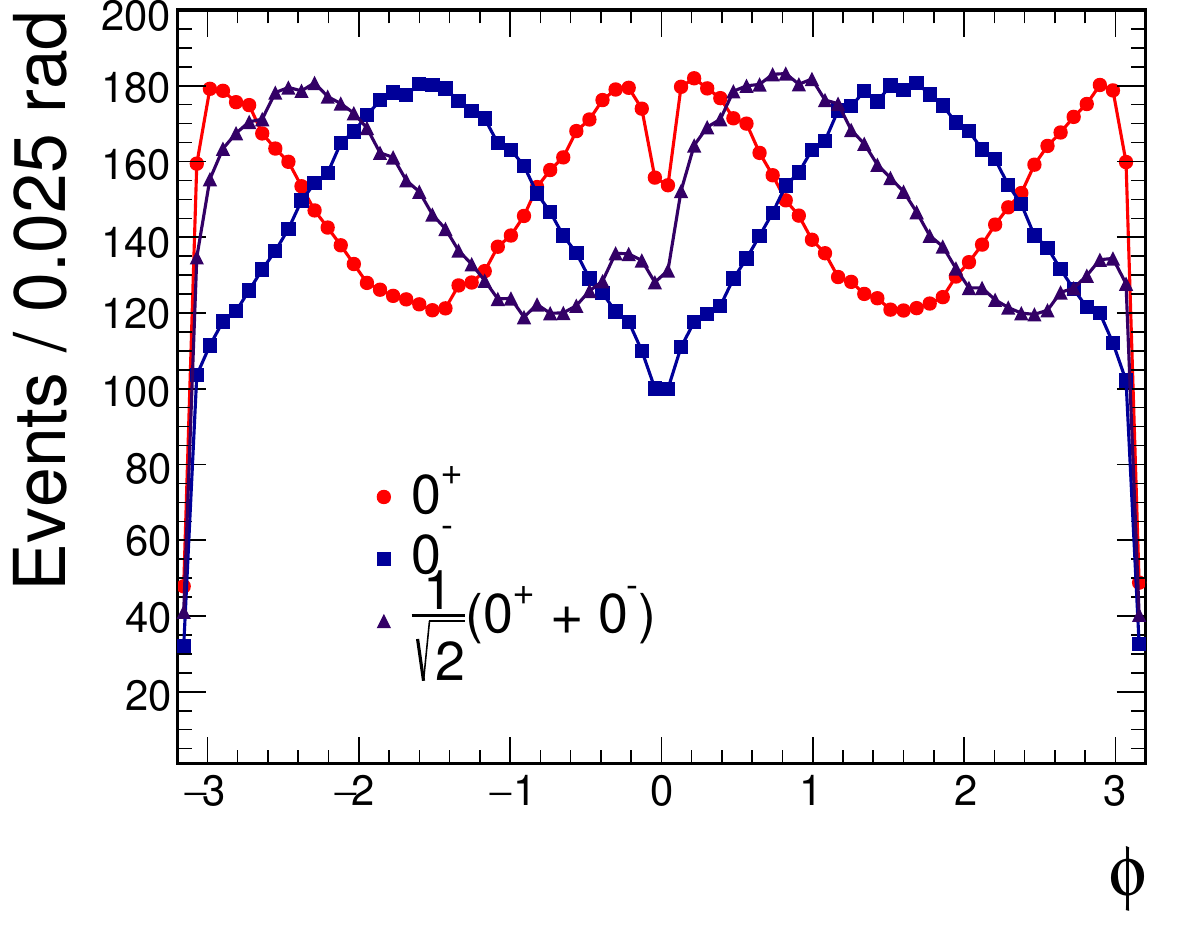}\
    \end{subfigure}
    \begin{subfigure}[t]{0.48\textwidth}
    \includegraphics[width = \textwidth]{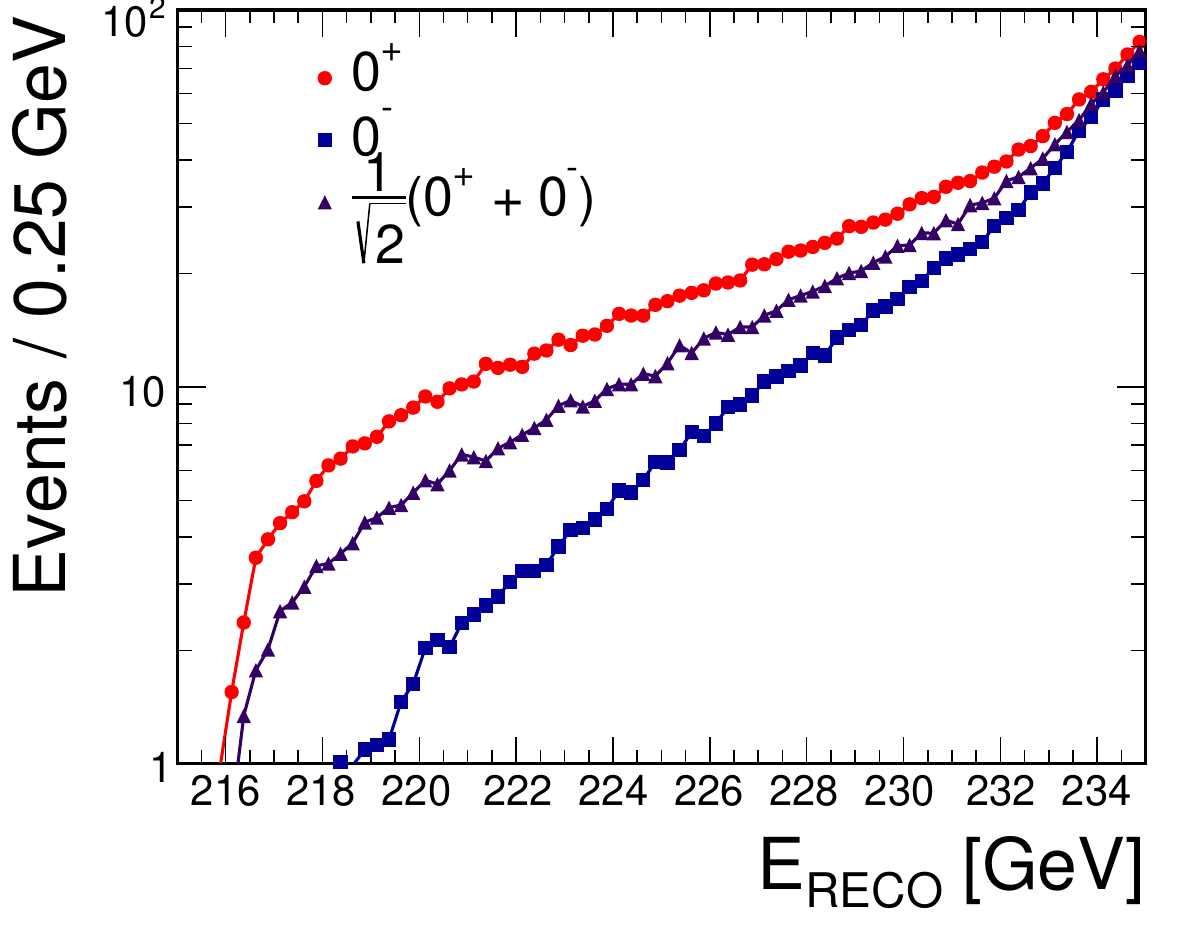}\
    \end{subfigure}

    \begin{subfigure}[t]{0.48\textwidth}
    \includegraphics[width = \textwidth]{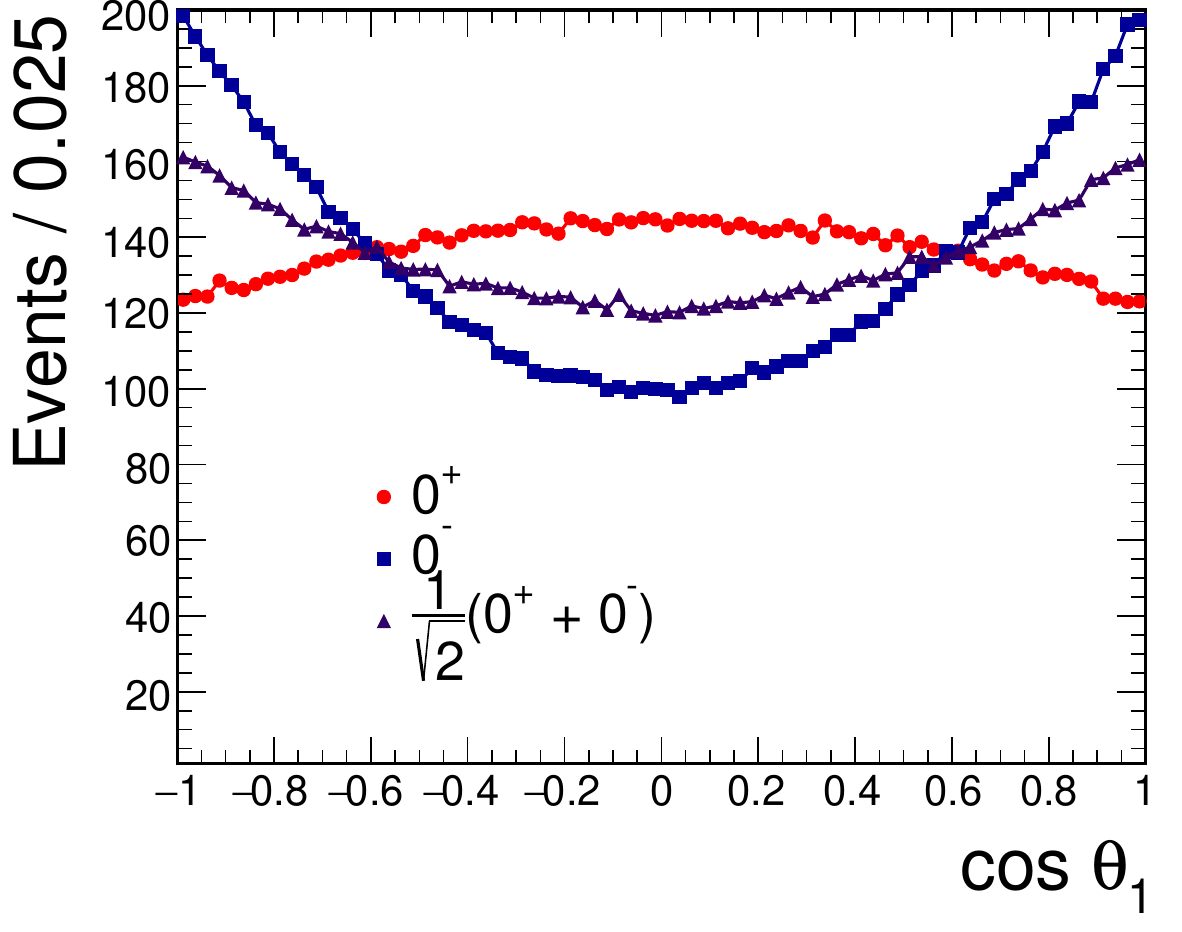}\
    \end{subfigure}
    \begin{subfigure}[t]{0.48\textwidth}
    \includegraphics[width = \textwidth]{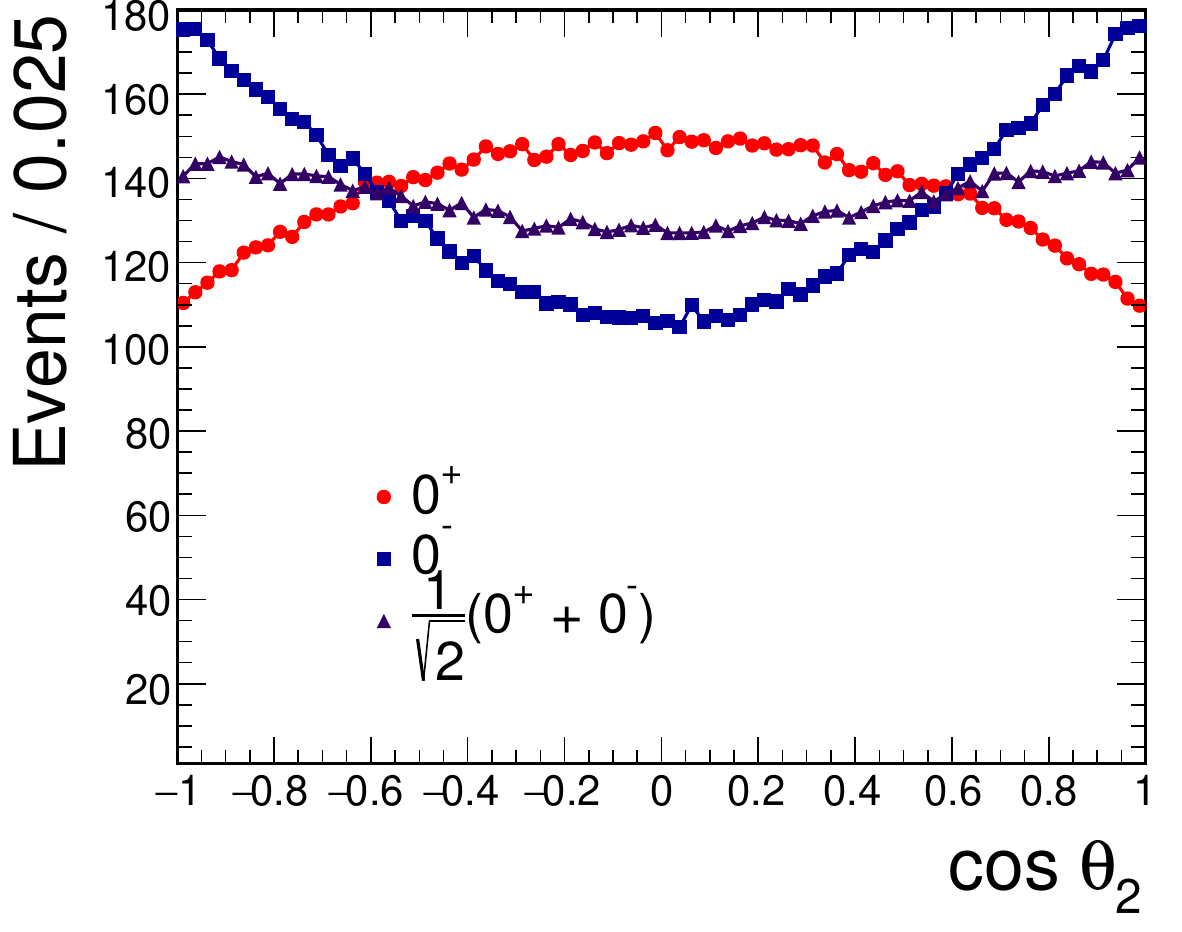}
    \end{subfigure}
    \caption{Distributions of the observables for the different Higgs boson $CP$ properties in the $e^+e^-\to HZ,~Z \to \mu^+\mu^-$ process at $\sqrt{s}=240~\text{GeV}$. 
    The $\phi$ (top-left picture), $E_{\text{RECO}}$ (top-right picture), $\cos \theta_1$ (bottom-left), $\cos \theta_2$ (bottom-right) distributions are shown.
    The samples with different Higgs boson $CP$ properties are shown: red circles for $0^+$, blue squares for $0^-$ and purple triangles for $1/\sqrt{2}\left( \ket{0^+} + \ket{0^-}\right)$ quantum numbers.}
    \label{fig:benchmark_samples}
\end{figure*}

Three benchmark samples are generated to illustrate the difference in the angular distributions 
for the different Higgs boson $CP$ properties:  
a pure $CP$-even scalar state $\ket{0^+}$,  
a pure $CP$-odd pseudoscalar state $\ket{0^-}$,  
and a mixed state $1/\sqrt{2}\left( \ket{0^+} + \ket{0^-}\right)$.  
The distributions of $\phi$, $\theta_1$, and $\theta_2$ are shown in Fig.~\ref{fig:benchmark_samples}.


\subsection{ISR energy shift}

In addition to the angular observables, another variable sensitive to the Higgs 
$CP$ properties is examined.  
ISR photons are predominantly emitted close to the beam axis, 
carrying away part of the event energy.  
The reconstructed total event energy, corrected for the ISR effects, is defined as
\begin{equation}
E_{\text{RECO}} = E_H + E_Z.
\end{equation}
Since the Higgs boson is not directly reconstructed, $E_H$ is obtained under the assumption of zero total event momentum:
\begin{equation}
E_H = \sqrt{p_H^2+m_H^2} \approx \sqrt{p_Z^2+m_H^2}.
\end{equation}
The $E_{\text{RECO}}$ distribution depends on the ISR energy spectrum and the cross section $\sigma(\sqrt{s})$ of the process. 
\begin{multline}
\sigma(e^+e^-\to HZ, \sqrt{s}, E_{\text{ISR}}) \approx \\ \approx \sigma(e^+e^-\to HZ, E_{\text{RECO}}, 0).
\end{multline}
While the ISR spectrum does not depend on the $CP$ properties of the process, the cross section does.
The ratio of the experimental to theoretical distributions for 
the $E_{\text{RECO}}$ variable  corresponds to the ratio of the experimental to 
theoretical Born cross sections for the process $e^+e^-  \to HZ$. Therefore, the 
measurement of the  $E_{\text{RECO}}$ distribution is effectively equivalent to the energy scan in 
the region between the sum of the Z and Higgs boson masses and $\sqrt{s}$. This can potentially be 
used for tests of other BSM models.

\begin{figure}[h]
    \centering
    \includegraphics[width=0.48\textwidth]{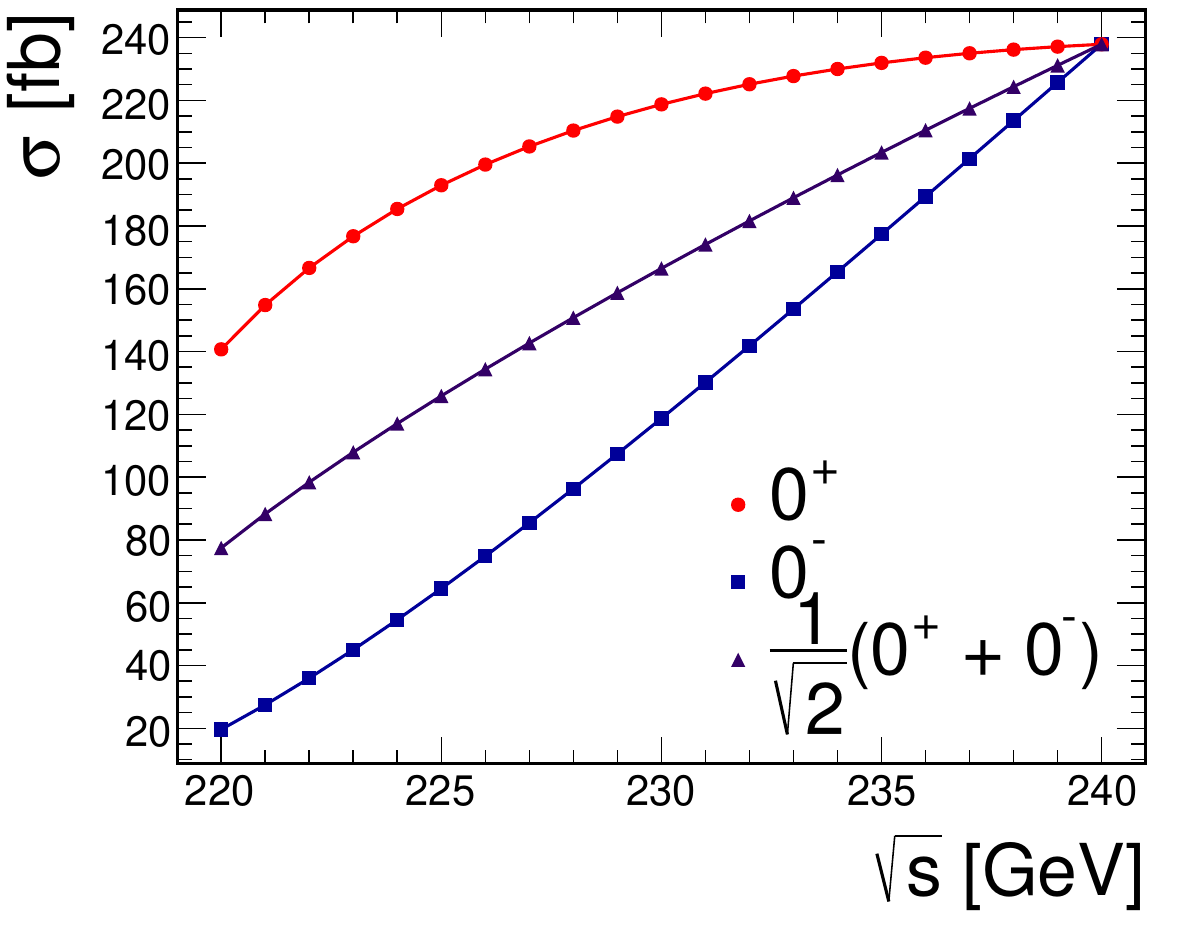}
    \caption{Dependence of the cross section $\sigma(\sqrt{s})$ on the center of mass energy for different Higgs boson $CP$ properties in the $e^+e^-\to HZ$ process.
    The red circles correspond to $0^+$ sample, blue squares to $0^-$, purple triangles to $1/\sqrt{2}\left( \ket{0^+} + \ket{0^-}\right)$ admixture.
    The equal cross section is assumed at $\sqrt{s}=240~\text{GeV}$.}
    \label{fig:higgs_xsec}
\end{figure}

The process $e^+e^- \to ZH$ proceeds through the $s$-channel with an intermediate $Z^*$ boson.  
For arbitrary Higgs boson $CP$ properties, the amplitude can be expressed as
\begin{equation}
M = \cos\psi_{CP}\,M^+ + \sin\psi_{CP}\,M^-,
\end{equation}
where $M^+$ and $M^-$ correspond to the transitions $1^- \to 0^+1^-$ and $1^- \to 0^-1^-$, respectively.  
In $M^+$, the final-state particles are produced in an $s$-wave, while in $M^-$ they are produced in a $p$-wave, resulting in different cross section behaviors.  
The dependence of $\sigma(\sqrt{s})$ on the center of mass energy for the scalar $\ket{0^+}$, 
pseudoscalar $\ket{0^-}$ and mixed Higgs boson are shown in Fig.~\ref{fig:higgs_xsec}. The cross 
sections are calculated assuming no ISR effects. 

The difference in the cross sections shapes leads to variations in the $E_{\text{RECO}}$ distributions.  
The distributions for the three benchmark samples described in the previous subsection are shown in 
Fig.~\ref{fig:benchmark_samples} (top-right picture).  
This observable provides complementary sensitivity and can be used to improve upper limits on the 
$CP$-odd admixture in the Higgs sector.

\begin{figure}[h!]
    \includegraphics[width = 0.48\textwidth]{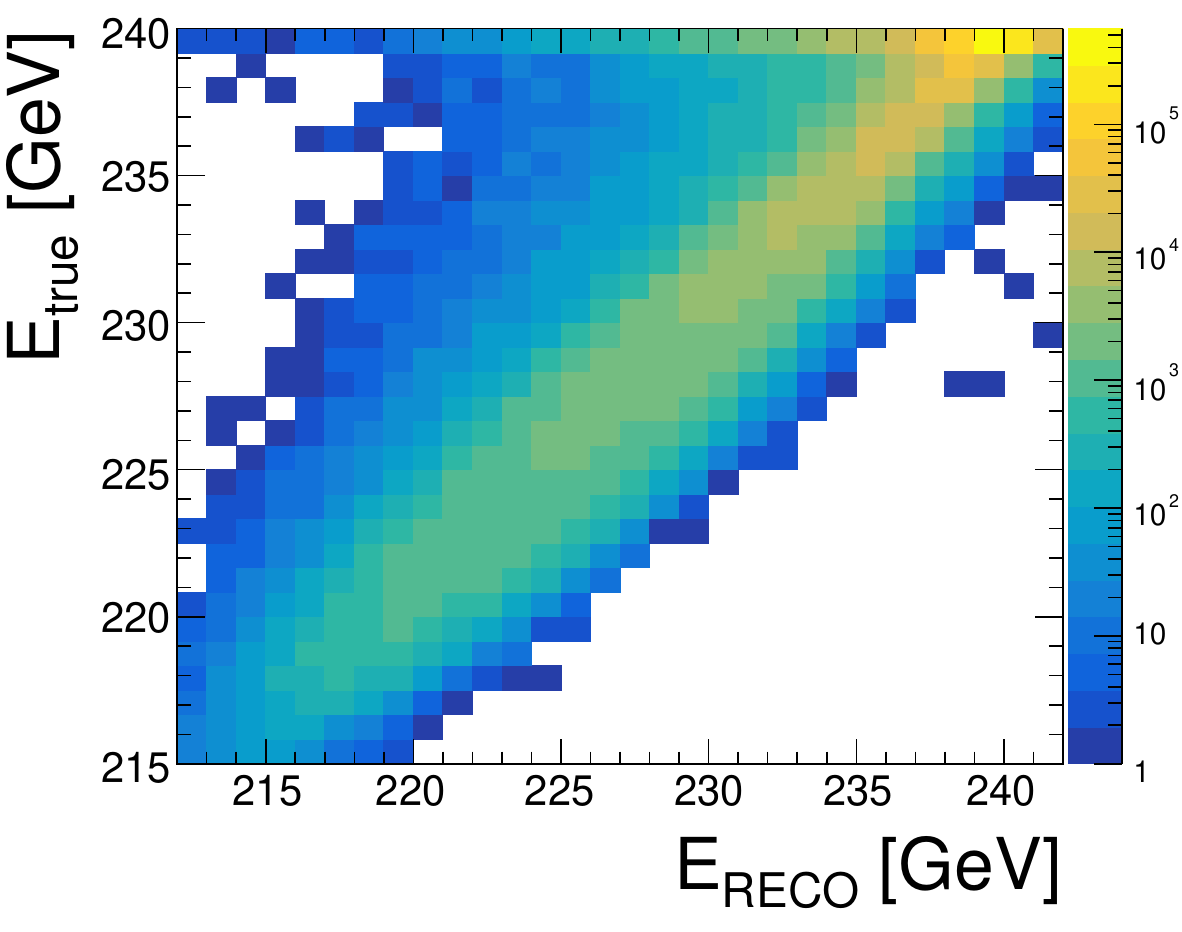}
    \caption{$E_{\text{true}}$ vs $E_{\text{RECO}}$ for the signal process $e^+e^-\to HZ$ after dressing muons and applying preselection on $M_{\mu^+\mu^-}$.}
    \label{fig:2d_resolution}
\end{figure}

\begin{figure}[h]
    \centering
    \includegraphics[width=0.48\textwidth]{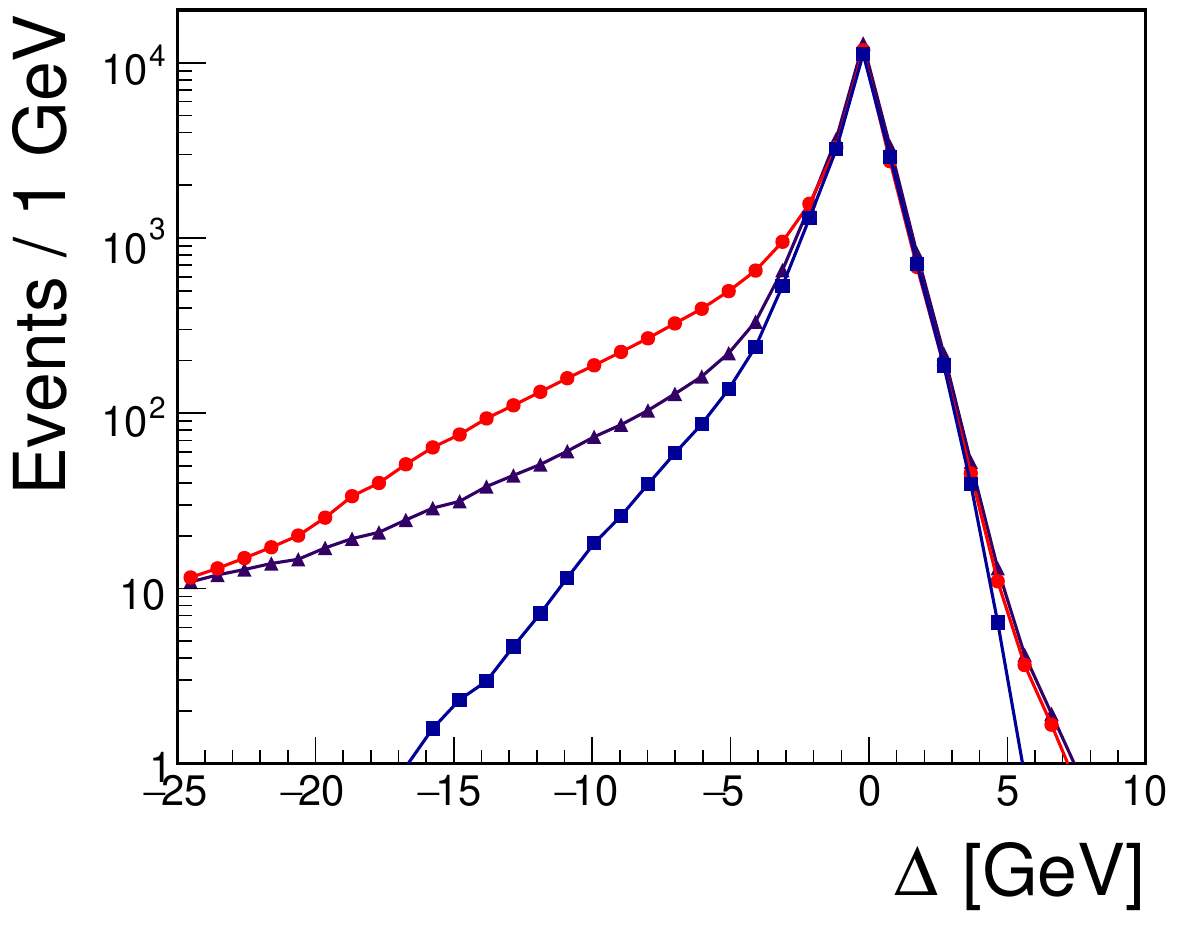}
    \caption{$\Delta = E_{\text{true}}-E_{\text{RECO}}$ distribution before (red circles), after dressing muons (purple triangles), after dressing muons and applying preselection on $M_{\mu^+\mu^-}$ (blue squares).}
    \label{fig:1d_resolution}
\end{figure}

To use this observable in the analysis, the event energy has to be accurately reconstructed. However, 
due to the FSR effects, the distribution of 
the difference between the true and reconstructed total event energies $\Delta$ 
exhibits a sizable non-Gaussian tail.
\begin{equation}
\Delta = E_{\text{true}} - E_{\text{RECO}},
\end{equation}
where $E_{\text{true}}$ is the total event energy obtained from the MC true collection. To mitigate these effects and improve the 
energy resolution, we dress the muons with FSR photons reconstructed in the detector within a cone of $\Delta R = \sqrt{(\Delta \eta^2 + \Delta \phi^2)} < 0.3$ around the muons and apply a preselection on the dimuon invariant 
mass, $M_{\mu^+\mu^-} > 86~\text{GeV}$. Figure~\ref{fig:2d_resolution} shows the 
two dimensional distribution of $E_{\text{true}}$ 
versus $E_{\text{RECO}}$ after the preselection. Figure~\ref{fig:1d_resolution} shows 
the one dimensional distribution of $\Delta$ before and after applying the 
preselection and dressing the muons. The bin content is shown in the logarithmic scale.
After the preselection procedures the effective Gaussian width for the variable
$\Delta$ is obtained to be $0.48~\text{GeV}$. 

The ISR assisted observable is directly sensitive to the absolute
muon momentum scale. The CEPC Reference Detector TDR requires a per mille level
transverse momentum resolution for leptons with momenta below
$100~\mathrm{GeV}$~\cite{cepc_ref_tdr}. Using the combined tracker
parameterization given in the TDR, the expected resolution for a
barrel muon with $p_T\simeq50~\mathrm{GeV}$, which is representative
of the muons considered in this analysis, is of the order of
$\sigma_{p_T}\simeq0.1~\mathrm{GeV}$. This event by event momentum
resolution should be distinguished from the uncertainty in the
absolute momentum scale, which can be determined much more precisely
using large calibration samples.

The Bhabha events, together with
the precise track information provided by the inner and outer silicon
trackers, can be used to determine and correct spatial distortions in
the TPC. The residual absolute momentum scale for muons can be
monitored using high statistics $e^+e^- \to \mu^+\mu^-$ and $Z\to\mu^+\mu^-$ samples and
the precisely known $Z$ boson mass. The large calibration statistics
expected at the CEPC should therefore allow the uncertainty in the
muon momentum scale to be kept well below the single muon momentum
resolution.

Consequently, the contribution of the absolute muon momentum scale
uncertainty to $E_{\mathrm{RECO}}$ is expected to be much smaller than the
effective $0.48~\mathrm{GeV}$ resolution of the $\Delta$ obtained in this analysis.
Nevertheless, the residual momentum scale uncertainty may induce
event migrations near the boundaries of the $E_{\mathrm{RECO}}$
intervals and should be propagated as an experimental systematic
uncertainty in a future analysis of collision data.

\begin{figure*}
    \begin{subfigure}[t]{0.48\textwidth}
    \includegraphics[width = \textwidth]{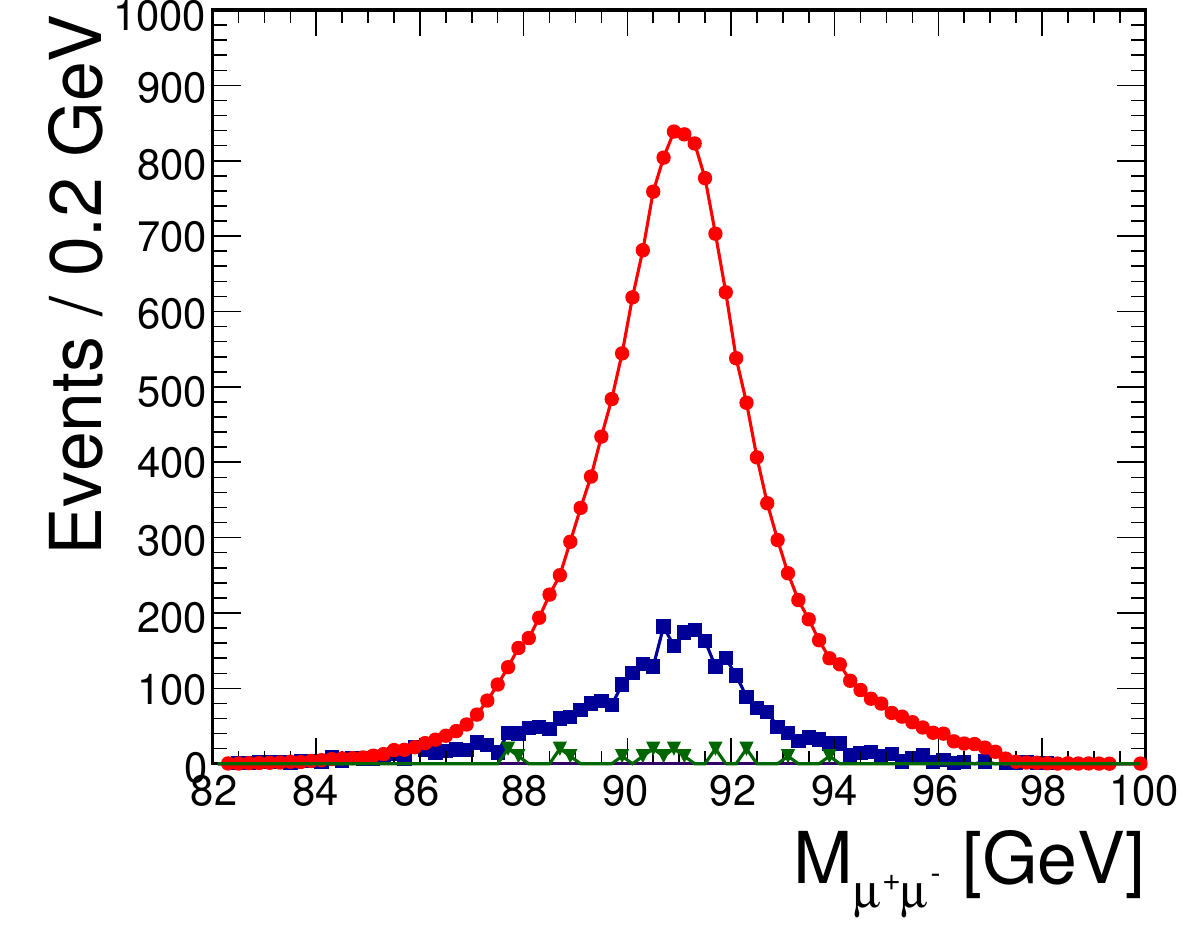}
    \end{subfigure}
    \begin{subfigure}[t]{0.48\textwidth}
    \includegraphics[width = \textwidth]{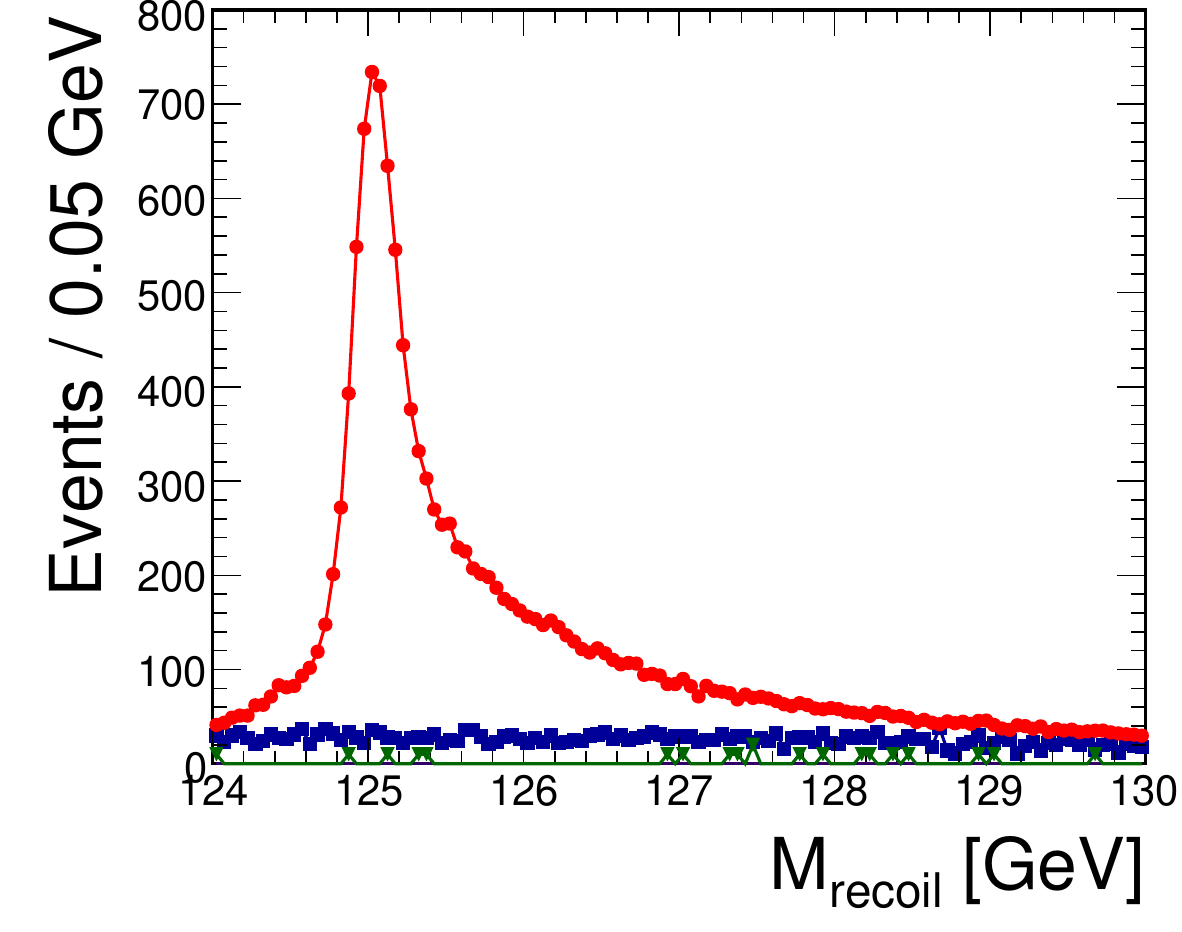}
    \end{subfigure}

    \begin{subfigure}[t]{0.48\textwidth}
    \includegraphics[width = \textwidth]{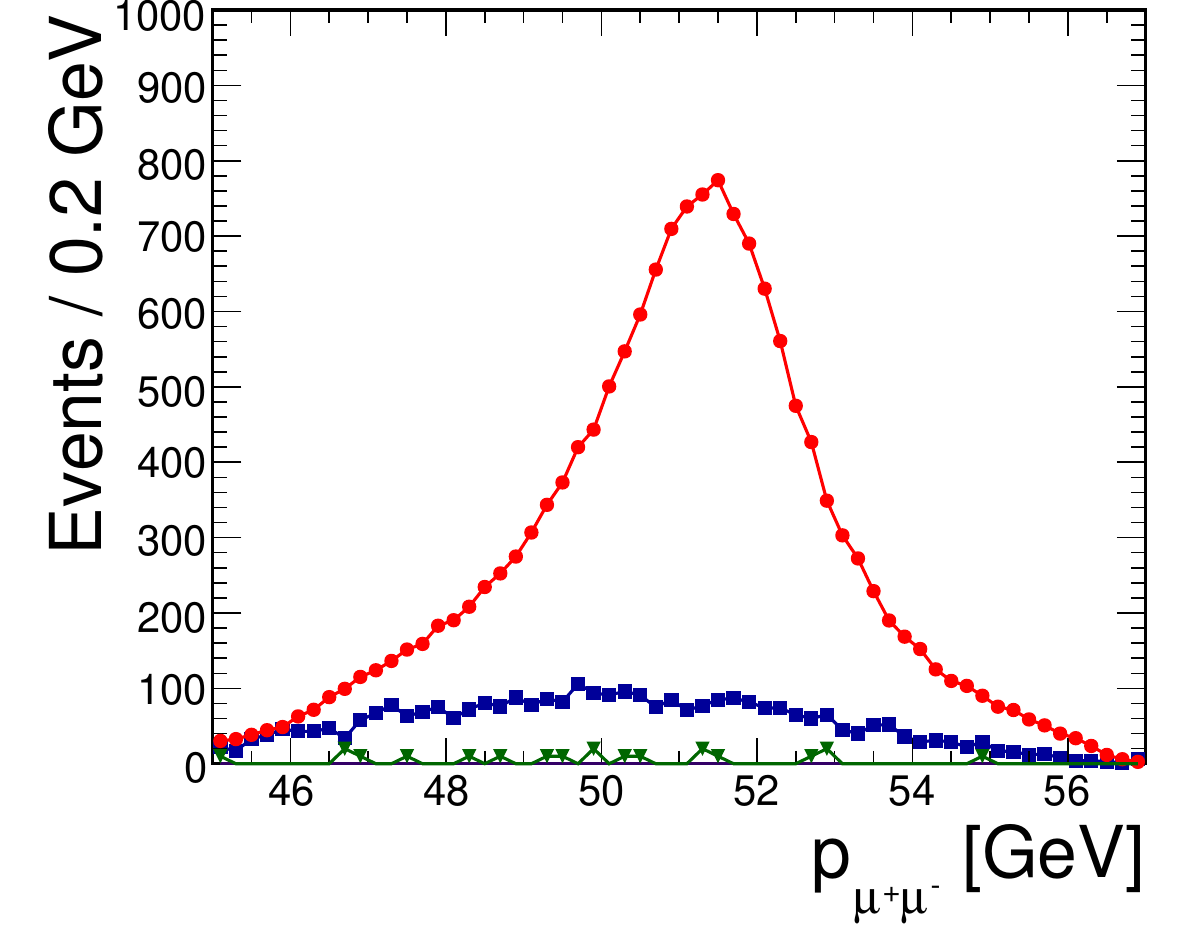}
    \end{subfigure}
    \begin{subfigure}[t]{0.48\textwidth}
    \includegraphics[width = \textwidth]{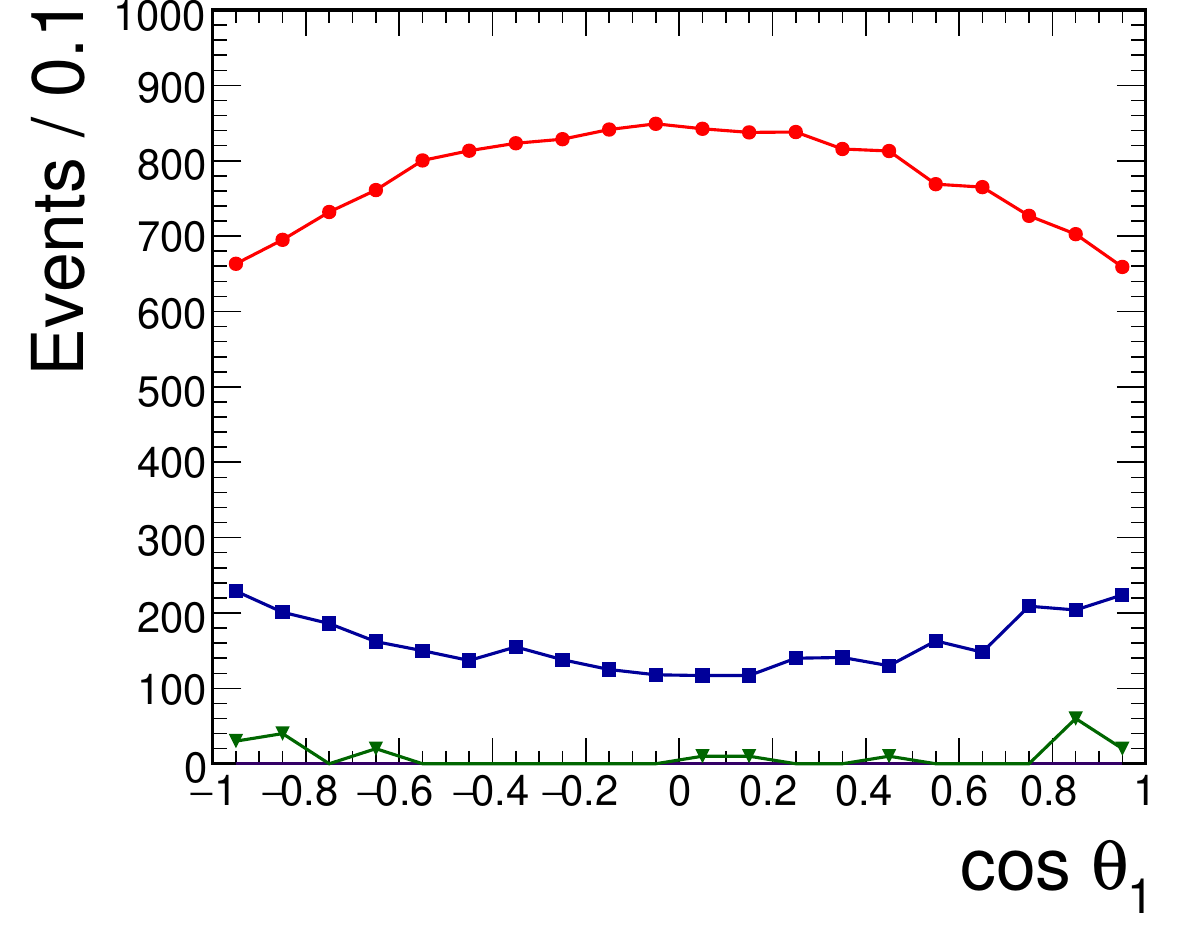}
    \end{subfigure}
    \caption{Distributions of the observables in the signal and background processes at $\sqrt{s}=240~\text{GeV}$. 
    The $M_{\mu^+\mu^-}$ (top-left picture), $M_{\text{recoil}}$ (top-right picture), $p_{\mu^+\mu^-}$ (bottom-left) and $\cos \theta_1$ (bottom-right) distributions are shown.
    The $HZ$ (signal) process is represented by red circles, $ZZ$ by blue squares, $WW/ZZ$ by purple up-triangles and $\mu^+\mu^-$ by green down-triangles.}
    \label{fig:backgrounds}
\end{figure*}

\subsection{Backgrounds and selections}

The main background processes for the studied signal are generated, including $\mu^+\mu^-$ process and 4-fermion processes, categorised as
$ZZ$ ($\mu^+\mu^- + q\bar{q}/e^+e^-/\nu_e\bar{\nu_e}/\nu_\tau\bar{\nu_\tau}$) and $WW/ZZ$ ($\mu^+\mu^-\nu_\mu\bar{\nu_\mu}$) with intereference included:
\begin{enumerate}[itemsep=-0.1cm]
    \item $e^+e^- \to \mu^+\mu^-$;
    \item $e^+e^- \to ZZ$;
    \item $e^+e^- \to WW/ZZ$.
\end{enumerate}

We include in calculations the leading order diagrams with on-shell resonant $Z$ and $W$ production and also 
subleading order nonresonant diagrams. The subleading diagrams give a few percent contribution in the cross sections.

To suppress these backgrounds, additional event preselection requires two oppositely 
charged muons accompanied by additional detector activity: at least one reconstructed jet or identified electron.  
This preselection effectively suppresses the $\mu^+\mu^-$ and $WW/ZZ$ processes, although a small 
number of events pass the preselection due to FSR and hadronization effects. The event numbers and
preselections efficiencies are given in Table~\ref{tab:preselections}.

\begin{table}[h!]
    \caption{Numbers of events in the studied samples for the preselection described above and $\mathcal{L}_{\rm int} = 5.6~\text{ab}^{-1}$,  $\sqrt{s} = 240~\text{GeV}$.}
    \begin{center}
        \begin{ruledtabular}
            \begin{tabular}{l c c c}
                Process & Events before & Events after & $\varepsilon~(\%)$ \\
                $e^+e^- \rightarrow$ & preselection & preselection & \\ \hline 
                $HZ,\quad Z\to\mu^+\mu^-$ & 37832 & 27453 & $72.56$ \\ 
                $ZZ$ & 2205920 & 1000920 & $45.37$ \\ 
                $\mu^+\mu^-$ & 31879327 & 30821 & $0.09$ \\ 
                $WW/ZZ$ & 1237982 & 293 & $0.02$ \\ 
            \end{tabular}
        \end{ruledtabular}
    \end{center}
    \label{tab:preselections}
\end{table}

A residual contribution from $ZZ$ and $\mu^+\mu^-$ events remains after the preselection and is further 
suppressed by different requirements applied in different $E_{\text{RECO}}$ ranges.

The following observables are considered to suppress the background events:
\begin{enumerate}[itemsep=-0.1cm]
    \item $M_{\mu^+\mu^-}$ --- the invariant mass of the two muon system; mainly suppresses the $e^+e^-\to\mu^+\mu^-$ events and FSR effects,
    \item $M_{\text{recoil}}$ --- the recoil mass to the two muon system; mainly suppresses the $e^+e^-\to ZZ$ events,
    \item $p_{\mu^+\mu^-}$ --- the momentum of the two muon system,
    \item $\cos \theta_1$ --- the angle of the $HZ$ production, suppresses the $e^+e^-\to ZZ$ events.
\end{enumerate}

The observables are investigated in four $E_{\text{RECO}}$ ranges 
and the following optimal $p_{\mu^+\mu^-}$ requirements are chosen for different 
$E_{\text{RECO}}$ ranges:

\begin{enumerate}
    \item $238 < E_{\text{RECO}} < 242~\text{GeV}$, 
    $45 < p_{\mu^+\mu^-} < 57~\text{GeV}$;

    \item $235 < E_{\text{RECO}} < 238~\text{GeV}$, 
    $42 < p_{\mu^+\mu^-} < 54~\text{GeV}$;

    \item $230 < E_{\text{RECO}} < 235~\text{GeV}$, 
    $39 < p_{\mu^+\mu^-} < 49~\text{GeV}$;

    \item $215 < E_{\text{RECO}} < 230~\text{GeV}$, 
    $26 < p_{\mu^+\mu^-} < 42~\text{GeV}$.
\end{enumerate}

This approach suppresses background processes and accounts for the evolution of the angular observables with $E_{\text{RECO}}$.
For all ranges the cuts $86 < M_{\mu^+\mu^-} < 96~\text{GeV}$ and $|\cos \theta_1 | < 0.98$ are applied. 
The remaining numbers of events for each process in each $E_{\text{RECO}}$ 
range after the selections are given in Table~\ref{tab:selections}. 
The distributions for these observables in the first $E_{\text{RECO}}$ range for the signal and background processes are shown in Fig.~\ref{fig:backgrounds}.
The figures~\ref{fig:backgrounds} are obtained with all the cuts applied except the cut on the variable shown.

\begin{table}[h!]
    \caption{Numbers of events in the studied samples for the selections described above in different $E_{\text{RECO}}$ ranges.}
    \begin{center}
        \begin{ruledtabular}
            \begin{tabular}{l c c c c}
                Process/Range [GeV] & $[238,242]$ & $[235,238]$ & $[230,235]$ & $[215,230]$ \\ \hline 
                $HZ$ & 14901 & 2309 & 3152  & 400\\ 
                $ZZ$ & 2687 & 832 & 3372 & 289\\ 
                $\mu^+\mu^-$ & 12 & 18 & 18 & 10\\ 
                $WW/ZZ$ & 2 & 2 & 1 & 3 \\ 
            \end{tabular}
        \end{ruledtabular}
    \end{center}
    \label{tab:selections}
\end{table}

\section{Analysis}

The analysis is based on a three-dimensional, binned likelihood constructed from the angular observables $\phi$, $\cos\theta_1$, and $\cos\theta_2$:
\begin{equation}
\mathcal{L} = \prod_{i,j,k} \text{Poiss}(\mu_{i,j,k}\,|\,N_{i,j,k}),
\end{equation}
where $i$, $j$, and $k$ denote the bin indices in $\phi$, $\cos\theta_1$, and $\cos\theta_2$, respectively. Here, $N_{i,j,k}$ represents the observed 
number of events in each bin, and $\mu_{i,j,k}$ is the expected number of events according to the SM. 
Since detector effects smear the 
observables, no analytical model is available for $\mu_{i,j,k}$. Instead, the expected $\mu_{i,j,k}$ values are 
obtained from a simulated sample containing 
approximately two orders of magnitude more events than the analysis sample, followed by rescaling to the same 
luminosity. This procedure reduces 
the statistical uncertainty in the $\mu$ distribution. The likelihood is evaluated in four $E_{\text{RECO}}$ 
intervals, each with the respective selection 
criteria.

\begin{figure}[t]
    \centering
    \includegraphics[width=0.48\textwidth]{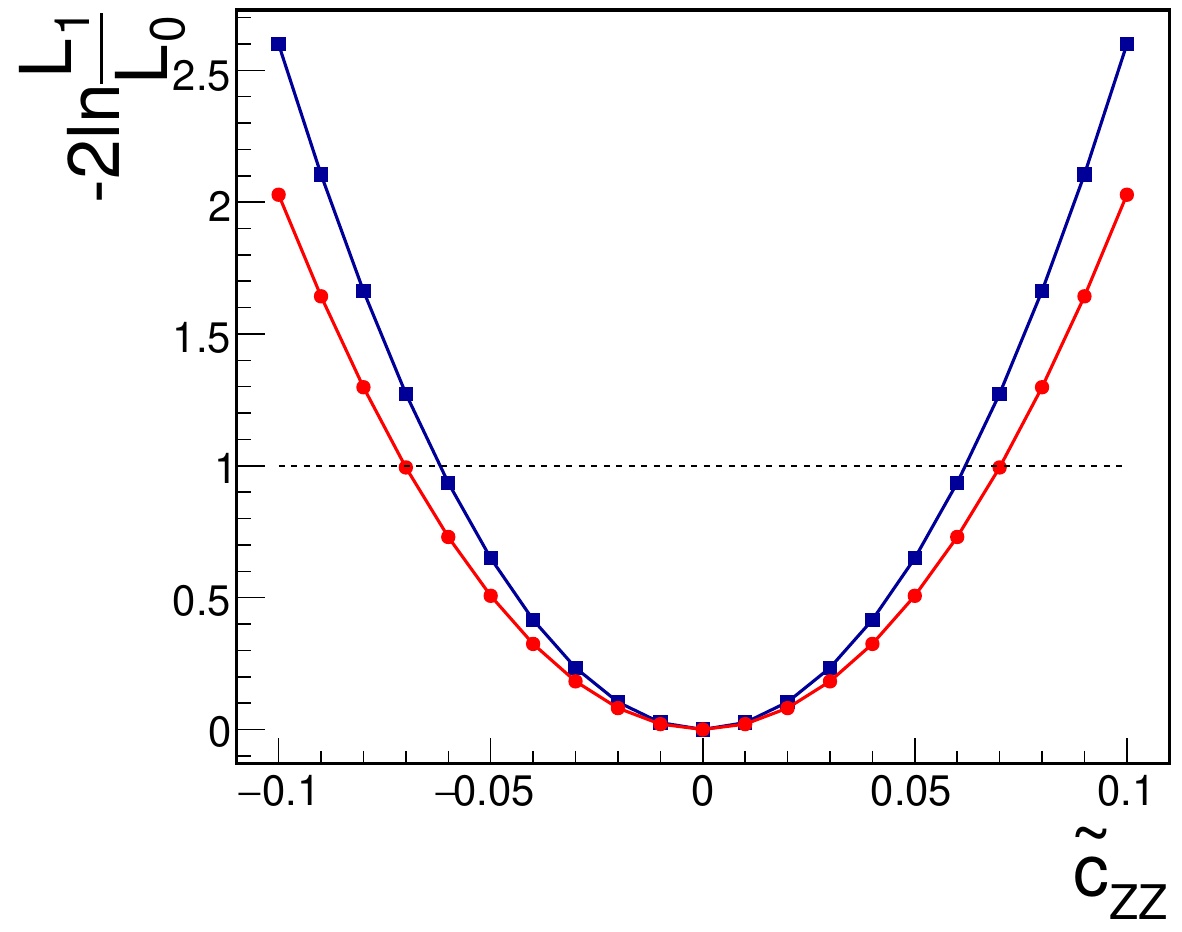}
    \caption{Dependence of the test statistic $-2\ln(\mathcal{L}_1/\mathcal{L}_0)$ on $\widetilde{c}_{ZZ}$. The points correspond to simulated 
    samples and the curve shows the fit with $a x^2 + b x^4$.
    The red line with circles corresponds to the angular likelihood 
    analysis, blue line with squares corresponds to ISR corrected analysis. 
    Dashed line shows the $68\%$ CL($1\sigma$).}
    \label{fig:significance_fit}
\end{figure}

For comparison, an additional analysis is performed without applying ISR corrections in order to evaluate the improvement in 
the upper limit obtained with the proposed method. The likelihood construction is identical, except that no binning in $E_{\text{RECO}}$ is 
applied. A single set of selection criteria, optimized for the best expected significance, is used for all events. The likelihood under the null 
hypothesis, $\mathcal{L}_0$, is computed for $\widetilde{c}_{ZZ}=0$ value, while the likelihoods under the alternative hypotheses, $\mathcal{L}_1$, 
are evaluated for simulated samples with $\widetilde{c}_{ZZ}$ values in the range $[-1.2, 1.2]$. The test statistic is defined as
\begin{equation}
S = -2\ln\frac{\mathcal{L}_1}{\mathcal{L}_0},
\end{equation}
and its dependence on $\widetilde{c}_{ZZ}$ is shown in Fig.~\ref{fig:significance_fit}. The obtained $S$ values are fitted with a polynomial function 
of the form $a x^2 + b x^4$. The fitting function is chosen for 
theoretical reasons. The total cross section of the process is 
proportional to the matrix element squared $|M|^2$, which 
can be decomposed in $CP$-even and $CP$-odd amplitudes as 
\begin{equation}
M = M^+ + \widetilde{c}_{ZZ}M^-,
\end{equation}
therefore the matrix element squared is a quadratic function of the $\widetilde{c}_{ZZ}$.
\begin{equation}
|M|^2 = |M^+|^2 + 2\widetilde{c}_{ZZ}\text{Re}(M^+(M^-)^{\dag}) + \widetilde{c}_{ZZ}^2|M^-|^2.
\end{equation}
This implies the form of the likelihood fitting function.
The result of the fit is shown in the Fig.~\ref{fig:significance_fit} both for analyses with $E_{\text{RECO}}$ correction and without. 
The upper limit from the analysis without the correction 
is consistent with the result~\cite{chinese_cp} for the $Z\to\mu^+\mu^-$ channel which is
$[-0.08, 0.07]$ for $\widetilde{c}_{ZZ}$ at the $68\%$ CL ($1\sigma$).
Including the $E_{\text{RECO}}$ correction improves the expected $68\%$ CL($1\sigma$) upper 
limit on $\widetilde{c}_{ZZ}$ from $0.071$ to $0.062$, corresponding to an improvement of approximately $15\%$.

\section{Conclusions}

In this work, we studied the sensitivity of the Circular Electron
Positron Collider experiment to possible $CP$-violating admixtures 
in the $HZZ$ coupling within the Higgs Characterisation framework. 
The Monte Carlo simulation including initial state radiation effects and detector response is performed 
for the process $e^+e^- \to ZH$ with $Z \to \mu^+\mu^-$ and relevant background processes. 
The main methodological result of this work is the development of a novel 
ISR-assisted data analysis method for probing the Higgs $CP$ structure. Rather than treating ISR solely as a radiative 
correction, the method exploits ISR induced variations in the reconstructed total event energy $E_{\text{RECO}}$, divides 
the data into effective collision energy intervals, and combines the corresponding energy dependent angular 
distributions in a multidimensional likelihood analysis.
This novel method improves the expected upper limit on the $CP$-odd coupling parameter 
$\widetilde{c}_{ZZ}$ from $0.071$ to $0.062$ at $68\%$ CL, corresponding to an improvement of approximately $15\%$ 
over the conventional analysis based solely on angular observables.
This demonstrates that ISR sensitive observables can provide complementary information and enhance the precision of $CP$ property 
measurements at future Higgs factories. The proposed method requires high muon energy resolution and low beam energy spread, and thus can be applied to 
other future Higgs factories such as FCC-ee, ILC, or CLIC without any additional detector modifications. 
Since polarised beams are expected at the ILC and CLIC colliders, the theoretical description of the observables would 
become more complex.
The developed approach can be extended to other Higgs production and decay channels, offering a promising path toward comprehensive 
and model independent tests of $CP$ symmetry in the Higgs sector at next generation lepton colliders.

\section*{Acknowledgments}

The authors are grateful to Li Gang for providing comments on the paper.

\section*{Funding}

This work is supported by the Russian Science Foundation grant number 25-22-00716.

\section*{Conflict of interest}

The authors of this work declare that they have no conflicts of interest.


\begin{thebibliography}{26}

    \bibitem{201230}
    S.~Chatrchyan,~V. Khachatryan, A.M.~Sirunyan et al. (CMS Collaboration), 
    "Observation of a new boson at a mass of 125 GeV with the CMS experiment at the LHC", \href{https://doi.org/10.1016/j.physletb.2012.08.021}{Physics Letters B {\bf 716}, 1 (2012).}

    \bibitem{2212.05833}
    G.~Aad, B.~Abbott, D.~C.~Abbott et al. (ATLAS Collaboration), 
    "Measurement of the CP properties of Higgs boson interactions with $\tau$-leptons with the ATLAS detector", 
    \href{https://doi.org/10.1140/epjc/s10052-023-11583-y}{Eur. Phys. J. C {\bf 83} (2023) 563.}

    \bibitem{cepc_whitemass}
    H.~Cheng, W.H.~Chiu, Y.~Fang et al., 
    "The Physics potential of the CEPC. Prepared for the US Snowmass Community Planning Exercise (Snowmass 2021)", \href{https://doi.org/10.48550/arXiv.2205.08553}{arXiv:2205.08553 [hep-ph].}

    \bibitem{cepc_a_tdr}
    W.~Abdallah, T.C.A.~de~Freitas, K.~Afanaciev et al.(The CEPC Study Group), 
    "CEPC Technical Design Report -- Accelerator (v2)", \href{https://doi.org/10.48550/arXiv.2312.14363}{arXiv:2312.14363 [physics.acc-ph]}

    \bibitem{cepc_ref_tdr}
    W.~Abdallah, T.C.A.~de~Freitas, K.~Afanaciev et al.(The CEPC Study Group), 
    "CEPC Technical Design Report -- Reference Detector", \href{https://doi.org/10.48550/arXiv.2510.05260}{arXiv:2510.05260 [hep-ex]}

    \bibitem{fcc}
    A.~Abada, M.~Abbrescia, S.~S.~AbdusSalam et al. (the FCC Collaboration),  
    "FCC-ee: The Lepton Collider", \href{https://doi.org/10.1140/epjst/e2019-900045-4}{Eur. Phys. J. Spec. Top. {\bf 228}, 261–623 (2019). }

    \bibitem{ilc_1}
    C.~Adolphsen, M.~Barone, B.~Barish et al., 
    "The International Linear Collider Technical Design Report - Volume 3.II: Accelerator Baseline Design", \href{https://doi.org/10.48550/arXiv.1306.6328}{arXiv:1306.6328 [physics.acc-ph].}

    \bibitem{ilc_2}
    T.~Barklow, J.~Brau, K.~Fujii, J.~Gao, J.~List, N.~Walker, K.~Yokoya, 
    "ILC Operating Scenarios", \href{https://doi.org/10.48550/arXiv.1506.07830}{arXiv:1506.07830 [hep-ex]}

    \bibitem{ilc_3}
    A.~Aryshev, T.~Behnke, M.~Berggren et al., 
    "The International Linear Collider: Report to Snowmass 2021", \href{https://doi.org/10.48550/arXiv.2203.07622}{arXiv:2203.07622 [physics.acc-ph]}

    \bibitem{clic_1}
    M.~Aicheler, P.N.~Burrows, N.~Catalan, R.~Corsini, M.~Draper, J.~Osborne, D.~Schulte, S.~Stapnes, M.J.~Stuart, 
    "The Compact Linear Collider (CLIC) – Project Implementation Plan", \href{https://doi.org/10.23731/CYRM-2018-004}{arXiv:1903.08655 [physics.acc-ph]}

    \bibitem{ilc_higgs}
    D.M.~Asner, T.~Barklow, C.~Calancha et al., 
    "ILC Higgs White Paper", \href{https://doi.org/10.48550/arXiv.1310.0763}{arXiv:1310.0763 [hep-ph].}

    \bibitem{1804.01241}
    D.~Jeans and G.~W.~Wilson, 
    "Measuring the $CP$-state of tau lepton pairs from Higgs decay at the ILC", \href{https://doi.org/10.1103/PhysRevD.98.013007}{Phys. Rev. D {\bf 98}, 013007 (2018).}

    \bibitem{chinese_cp}
    Q.~Sha, A.~Fadol, F.~Guo, G.~Li, Y.~Fang, J.~Gu and X.~Lou, 
    "Probing Higgs CP properties at the CEPC in the $e^+e^- \to ZH \to \ell^+\ell^-H$ using optimal variables", \href{https://doi.org/10.1140/epjc/s10052-022-10926-5}{Eur. Phys. J. C {\bf 82}, 981 (2022).}

    \bibitem{2304.04390}
    K.~Cheung, Y.-n.~Mao, S.~Moretti, R.~Zhang, 
    "Testing CP-violation in a heavy Higgs sector at CLIC", \href{https://doi.org/10.1140/epjc/s10052-025-14369-6}{Eur. Phys. J. C {\bf 85} (2025) 6, 700.}

    \bibitem{2412.13130}
    A.~Blondel, C.~Grojean, P.~Janot and G.~Wilkinson, 
    "Higgs Factory options for CERN: A comparative study", \href{https://doi.org/10.48550/arXiv.2412.13130}{arXiv:2412.13130 [hep-ph].}

    \bibitem{atlas_prl}
    G.~Aad, B.~Abbott, D.C.~Abbott et al. (ATLAS Collaboration), 
    "Combination of Measurements of $CP$ Properties of Higgs Boson Interactions with Vector Bosons Using Proton-Proton Collisions at $\sqrt{s}=13$ TeV with the ATLAS Detector", \href{https://doi.org/10.1103/54vv-lgc7}{Phys. Rev. Lett. {\bf 137}, 021801 (2026).}

    \bibitem{cms_cp}
    S.~Chatrchyan, V.~Khachatryan, A.M.~Sirunyan et al. (CMS Collaboration), 
    "Constraints on anomalous Higgs boson couplings to vector bosons and fermions using the $\gamma\gamma$ final state in proton-proton collisions at $\sqrt{s}=13$ TeV", \href{https://doi.org/10.48550/arXiv.2605.14245}{arXiv:2605.14245 [hep-ex].}

    \bibitem{ecfa}
    ECFA Collaboration, 
    "ECFA Higgs, electroweak, and top factory study", \href{https://doi.org/10.23731/CYRM-2025-005}{CERN-2025-005}

    \bibitem{edm_1}
    S.~Inoue, M.~J.~Ramsey-Musolf and Y.~Zhang, 
    "$CP$-violating phenomenology of flavor conserving two Higgs doublet models", \href{https://doi.org/10.1103/PhysRevD.89.115023}{Phys. Rev. D {\bf 89}, 115023 (2014).}

    \bibitem{edm_2}
    M.~Pospelov and A.~Ritz, 
    "Electric Dipole Moments and New Physics", \href{https://doi.org/10.48550/arXiv.2509.23531}{arXiv:2509.23531 [hep-ph].}

    \bibitem{hc}
    P.~Artoisenet, P.~de~Aquino, F.~Demartin, R.~Frederix, S.~Frixione, F.~Maltoni, M.~K.~Mandal, P.~Mathews, K.~Mawatari, V.~Ravindran, S.~Sethm ~P.~Torielli and M.~Zaro, 
    "A framework for Higgs characterisation", \href{https://doi.org/10.1007/JHEP11(2013)043}{J. High Energ. Phys. {\bf 2013}, 43 (2013). }

    \bibitem{atlas_hzz}
    G.~Aad, B.~Abbott, D.C.~Abbott et al. (ATLAS Collaboration), 
    "Prospects for measurements of tensor structure of the $HZZ$ vertex in $H\to ZZ^*\to 4\ell$ decay with ATLAS detector"
    \href{https://inspirehep.net/literature/1795425}{ATL-PHYS-PUB-2013-013}

    \bibitem{whizard_1}
    W.~Kilian, T.~Ohl, J.~Reuter, 
    "WHIZARD—simulating multi-particle processes at LHC and ILC", \href{https://doi.org/10.1140/epjc/s10052-011-1742-y}{Eur.Phys.J. {\bf C71} (2011) 1742.}
    
    \bibitem{whizard_2}
    M.~Moretti, T.~Ohl, J.~Reuter, 
    "O'Mega: An Optimizing Matrix Element Generator", \href{https://doi.org/10.48550/arXiv.hep-ph/0102195}{arXiv: hep-ph/0102195.}
    
    \bibitem{pythia}
    T.~Sjöstrand, L.~Lönnblad, and S.~Mrenna, 
    "PYTHIA 6.2 Physics and Manual", \href{https://doi.org/10.48550/arXiv.hep-ph/0108264}{arXiv:hep-ph/0108264}. 

    \bibitem{delphes}
    J.~de~Favereau, C.~Delaere, P.~Demin, A.~Giammanco, V.~Lemaître, A.~Mertens and M.~Selvaggi, 
    "DELPHES 3: a modular framework for fast simulation of a generic collider experiment", \href{https://doi.org/10.1007/JHEP02(2014)057}{J. High Energy Phys. 02 (2014) 057}.

\end{thebibliography}
\end{document}